\documentstyle[preprint,aps,pre,epsf]{revtex}     
\begin{document}
\draft
\tighten
\title{Microscopic Non-Universality versus Macroscopic Universality\\[6mm] 
in Algorithms for Critical Dynamics\\[17mm]}
\author{U. Ritschel\footnote{e-mail: uwe@theo-phys.uni-essen.de} and P. Czerner\footnote{e-mail: peterc@theo-phys.uni-essen.de}\\[13mm]}
\address{Fachbereich Physik, Universit\"at GH Essen, 45117 Essen
(F\ R\ Germany)\\[32mm]}
\narrowtext
\maketitle
\begin{abstract}
We study
relaxation processes in spin systems near criticality
after a quench from a high-temperature initial state.
Special attention is paid to the stage where universal behavior, with
increasing order parameter $m(t)\sim t^{\theta}$,
emerges from an early non-universal period.
We compare various algorithms, lattice types, and updating schemes
and find in each case the
same universal behavior at {\it macroscopic} times, despite of
surprising differences during the early non-universal stages.
\end{abstract}


Temperature quenches in spin systems
and the ensuing relaxational processes have
been a much-studied subject in the recent 
past\cite{bray,jans,jans2,huse,monte,liea,sczh,okano,oerd1,oerd2,oerd3,own,nucl,surf,grass,kothou,zheng,overshoot,maju}. 
Especially for a quench from
a high-temperature initial state to the critical region, a new
universal regime was predicted
by Janssen et al.\cite{jans}. In its most pronounced form this
phenomenon, termed universal short-time behavior (USTB) in the following,
occurs in a model with purely relaxational dynamics
(model A according to Ref.\,\cite{halp}).
Starting from an initial state with $T\gg T_c$ and a {\it small} initial
magnetization $m_0$, the magnetization increases as \cite{jans}
\begin{equation}\label{power}
m(t) \sim m_0\,t^{\theta}
\end{equation}
for a macroscopic time span
before it reaches a maximum and eventually
decays to the equilibrium value zero.
$\theta$ is an independent new exponent determined
by the non-equilibrium initial state 
that cannot be expressed in terms of equilibrium exponents.
Its value in the two-dimensional Ising system
for instance is $\theta\simeq 0.19$\cite{liea,grass}.

The results of Janssen et al.\cite{jans,jans2} allowed the interpretation
of earlier Monte Carlo (MC) simulations\cite{huse},
and later the power law (\ref{power}) was also directly verified
in the ``computer experiment''\cite{monte}. Further it was pointed out by
Li et al. \cite{liea} that the USTB can be exploited to 
determine {\it equilibrium} exponents from the early non-equilibrium.
In the sequel, this method was further
developed and applied
to a number of systems\cite{sczh,okano}. Moreover, 
USTB in other dynamic
universality classes\cite{oerd1}, in systems with a tricritical point
\cite{oerd2}, and in disordered and
dilute spin systems was studied\cite{oerd3}. 
Finite size effects were analyzed in Refs.\,\cite{own,nucl}, and 
the USTB near surfaces was studied in Ref.\,\cite{surf}.
Further, the
close relationship between USTB and ``damage spreading'' was pointed out
\cite{grass}, 
USTB in a different context, for
quasi-long-range order evolving after a quench to the Kosterlitz-Thouless
phase, was investigated \cite{kothou}, and
a general scaling invariance in the short-time regime
was found \cite{zheng}.
Possibly
also related to USTB is the ``overshooting'' of the order parameter
beyond the equilibrium domain magnetization for quenches
below $T_c$ \cite{overshoot}, and the issue whether
there exist still more independent exponents for relaxational
processes was raised in a recent preprint \cite{maju}. 

A simple physical argument for the growth
of the magnetization in (\ref{power})
was given by Janssen\cite{jans2,private}. 
Consider a system that is quenched to some
final temperature $T_f$ (not necessarily $T_c$), again
with initial magnetization $m_0$.
Then for $T_f\ll T_c$, $m(t)$
should grow after the quench, towards the equilibrium value
selected by $m_0$.
If in contrast $T_f\gg T_c$, 
$m(t)$ is expected to decay to zero rapidly. 
Hence, there should be a limiting temperature $T_l$
where the qualitative behavior changes.

As the initial correlations are short-ranged,
the natural candidate for $T_l$  
is the critical temperature
of the mean-field (MF) theory $T_c^{MF}$, and
with the real $T_c$ of spin systems
being always smaller than $T_c^{MF}$,
it would be an immediate consequence that $m(t)$
increases for a quench to the critical point. 
However, as argued in Ref.\,\cite{nucl}, it
is {\it not} possible to derive the power law (\ref{power}) from
this scenario. The power-law growth is rather a phenomenon that
occurs when the time-dependent (growing) correlation
length $\xi(t)$ has become {\it macroscopic},
i.e., much larger than the lattice spacing $a$
(compare Fig.\, 1 below). 
The derivation of (\ref{power})
is thus beyond the scope of MF theory.  

So far numerical 
investigations have been mostly carried out with the heat-bath (HB)
algorithm\cite{bind,foot1}.
A comparison between HB and Metropolis (ME)
algorithm \cite{bind} was performed for the Potts model
by Okano et al.\cite{okano}, and it turned out that
concerning the universal behavior both algorithms yield
compatible results, but differences occur for early times.

The main purpose of this Letter
is a more systematic examination of the issue of universality. 
Are equilibrium exponents determined with the 
USTB really independent of factors like the
algorithm (HB or ME),
the updating scheme (random or sequential),
and the lattice type (nearest or next-nearest neighbor coupling,
square or triangular lattice)?
How does universal behavior in the regime with $\xi(t)\gg a$
emerge from the non-universal early stage with $\xi(t)\simeq a$?
And closely related: Is it really a mean-field ordering
process during the {\it microscopically} early stages, or has
this simple picture to be refined?

We answered these questions by solving
the master equation for early times
(during the first single-spin updates) as well as by MC simulation
for later times.
Our work reveals a number of interesting and surprising details about
algorithms for critical dynamics and puts the USTB as a method
to access equilibrium properties from
a non-equilibrium (though universal)
regime on a much firmer basis. 

Information about
thermal averages in the early non-universal stage
of the relaxational process can be obtained by solving the
master equation \cite{bind}
\begin{equation}\label{master}
\frac{d P({\bf s},t)}{dt}=\sum_{{\bf s}'}\left[
W({\bf s}'\!\to\! {\bf s})\,P({\bf s}',t)-
W({\bf s}\!\to\! {\bf s}')\,P({\bf s},t)
\right]\>,
\end{equation}
where ${\bf s}$ denotes a spin configuration, $W({\bf s}'\!\to\!
{\bf s})$ is the
transition probability,
$P({\bf s},t)$ the
probability to find configuration ${\bf s}$ at time $t$, and
the sum extends over all possible configurations.
The analytic integration of Eq.\,(\ref{master}) for large systems
of coupled spins is not feasible.
However, for the high-temperature
initial state consisting of $N$ uncorrelated spins in a
magnetic field $H$,
characterized by the Boltzmann factor
\begin{equation}\label{hight}
P({\bf s},0)= Z_0^{-1} \exp(H\,\sum_{i=1}^N s_i)\quad \mbox{with}
\quad Z_0=(2 \cosh\,H)^{N}\>,
\end{equation}
the analytic treatment 
for very early times, $t\ll 1$,  is possible.
(The time is expressed in units of MC steps per site (MCS).)

Consider the first single-spin update after the quench.
It takes place in an environment that is coupled
to a heat bath at the {\it final} temperature $T_f$,
and $H$ is switched off.
Decisive for the very early stage is
whether after the {\it first} update on average the magnetization
is reduced or increased. The respective tendency survives
as long as the system
still closely resembles the initial state, i.e., as long as 
the number of single-spin updates is much smaller than $N$.

Without loss of generality one may choose spin 1 to be updated. 
From (\ref{master}) one
straightforwardly derives $\Delta m:
=m(t=1/N)-m_0$ to be
\begin{equation}\label{deltam1}
\Delta m =2\,N^{-1}\sum_{\bf s}\left[
W(-\!\to\! +)\,P(-,0)-W(+\!\to\! -)\,P(+,0)\right]\>,
\end{equation}
where the $W$'s are the probabilities for spin 1 changing sign
with all other spins remaining unchanged.
Especially for small $H$
(corresponding to small $m_0$), we find from (\ref{deltam1})
the simple result
\begin{equation}\label{deltam2}
\Delta m = -4\, H\,N^{-1}\sum_{\bf s} W(+\!\to\! -)\left(1+\sum_{i\in 
\mbox{\footnotesize IN}} s_i\right)\>,
\end{equation}
where the second sum extends over the interacting neighbors of spin 1. 

From (\ref{deltam2}) we
calculated $\Delta m$ and the limiting temperature
$T_l$ for the Ising model on a square lattice
with nearest-neighbor interaction $J$ for
HB and ME algorithm.
In $d=2$ (the system that will be also studied
by MC simulations below) the explicit results are $T_l^{HB}=
3.0898\ldots$ and $T_l^{ME}=1.5885\ldots$ for HB
and ME algorithm, respectively. 
(Temperatures are expressed in units of $J/k_B$.)  
For comparison, the (exact)
critical temperature is $T_c=2.2691\ldots$ and
$T_c^{MF}=4$.
Hence, for the HB algorithm $T_l$ is indeed above
$T_c$, while, surprisingly, with ME even {\it at} $T_c$
the magnetization decays $t\ll 1$.  

We calculated $T_l$ also for other dimensionalities.  
In the limit $d\to \infty$ (where the number of nearest
neighbors becomes infinite)
one finds $T_l\to T_c^{MF}$ for both algorithms considered above.
For $d=1$ both yield $T_l=0$. 

From this analysis we conclude:
First, the limiting temperature in general
depends on the algorithm. Only for
$d\to \infty$ it turns out to be $T_c^{MF}$.
Second, we are left with the puzzling result
that for the ME algorithm the limiting temperature lies even
{\it below} $T_c$, and therefore one would not expect to see
an increase of the magnetization at $T_c$. In any event, the
simple explanation that the non-universal stage preceding the USTB
is a MF ordering process is in general not correct. 
 
In order to learn about later stages, $t\gtrsim 1$, especially
the crossover from microscopic to macroscopic behavior, we had to
resort to MC simulations. These were carried out for
an Ising system on a square lattice with linear dimension
$L$ and periodically
coupled boundaries. Single spins were randomly selected and
updated. In order to obtain thermal
expectation values we generated a large number
of histories, each starting from
a new initial configuration, and calculated
mean values\cite{bind}.\\

\def\epsfsize#1#2{0.7#1}
\hspace*{2.5cm}\epsfbox{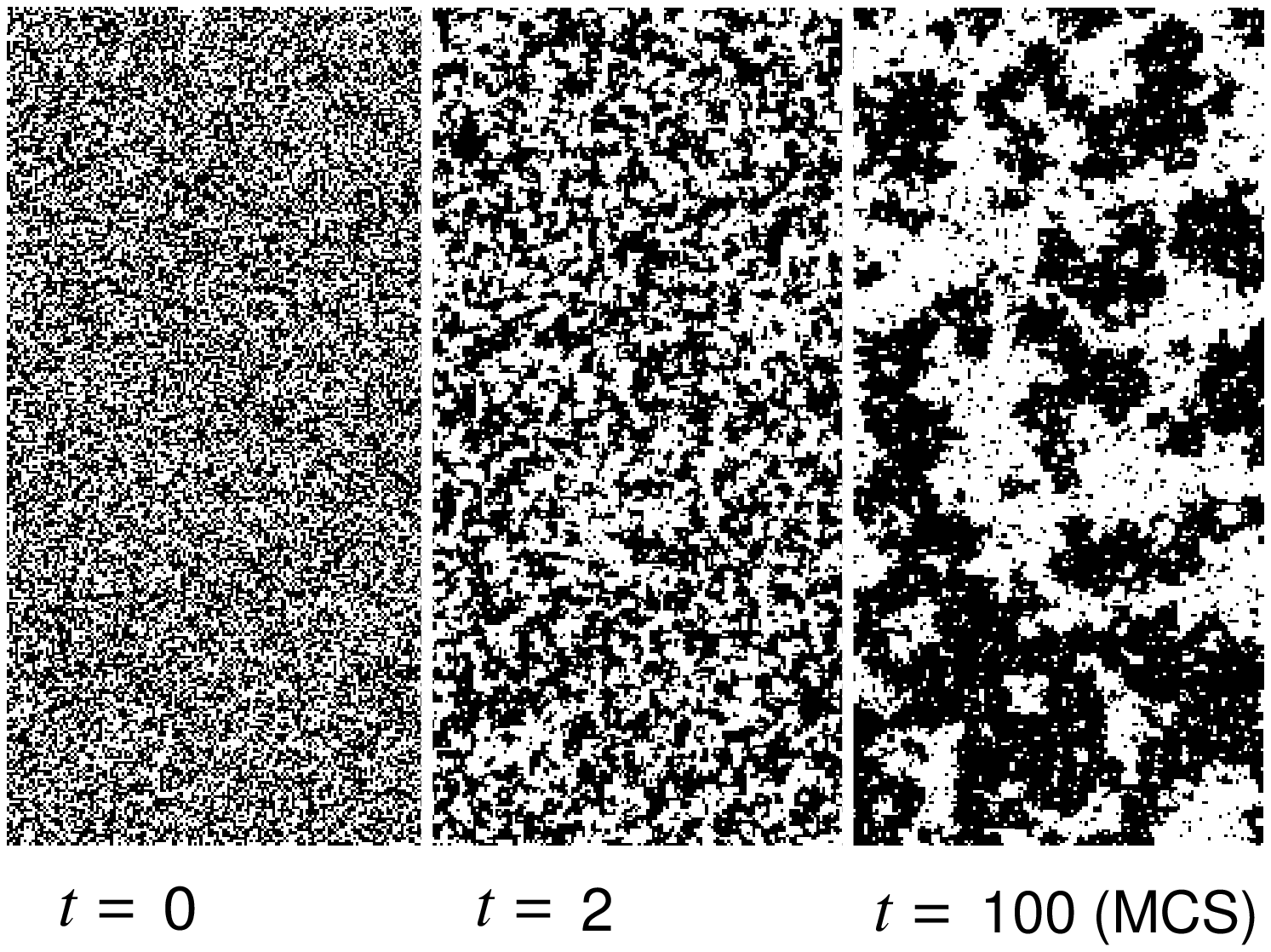}\\[0mm]
{\small {\bf Fig.\,1}: Three snapshots of the temporal evolution of a spin configuration for $L=300$ and $m_0=0$, generated with the
HB algorithm and random updating.
Displayed in each picture is half of the system. From the
visual appearance there is no difference HB and ME algorithm.} \\[0.1cm]

Snapshots of the temporal evolution of a single
configuration for a square lattice with $N=90\,000$ spins
are displayed in Fig.\,1. 
The left picture shows the initial state. Next to it
the configuration after $2\,N$ updates
corresponding to $t=2$ is depicted. 
At this point the average domain size
and (with that) the correlation length are already substantially larger than
the lattice spacing. This is the stage where universal
(macroscopic) behavior emerges from the non-universal (microscopic)
regime as discussed in more detail below.
At $t=100$ the correlation length is of the order
of the lattice size $L$.

Results for $m(t)$
at $T_f=T_c$ with $L=20$ and $m_0=0.05$
are displayed in Fig.\,2. 
The HB curve (solid line) monotonously increases
and is consistent with a power-law for $t\gtrsim 1.5$.
In the
case of the ME algorithm (dashed line) the behavior is
qualitatively different.
As expected from the our analytic results,
$m(t)$ indeed drops initially,
but has a minimum at $t\simeq 0.3$,
and then increases to assume the power-law
form for $t\gtrsim 2$.
Thus, despite of the anomalous time dependence of the ME
curve in the non-universal regime, for
macroscopic times it agrees with (\ref{power}).
This is in accord with the findings of Okano et al.,
where sequential updating
was used and, thus, the details of the temporal evolution were
not uncovered.
Later the profiles in Fig.\,2
have a maximum and then decay to the
equilibrium value zero\cite{foot2}.\\

\def\epsfsize#1#2{0.6#1}
\hspace*{2cm}\epsfbox{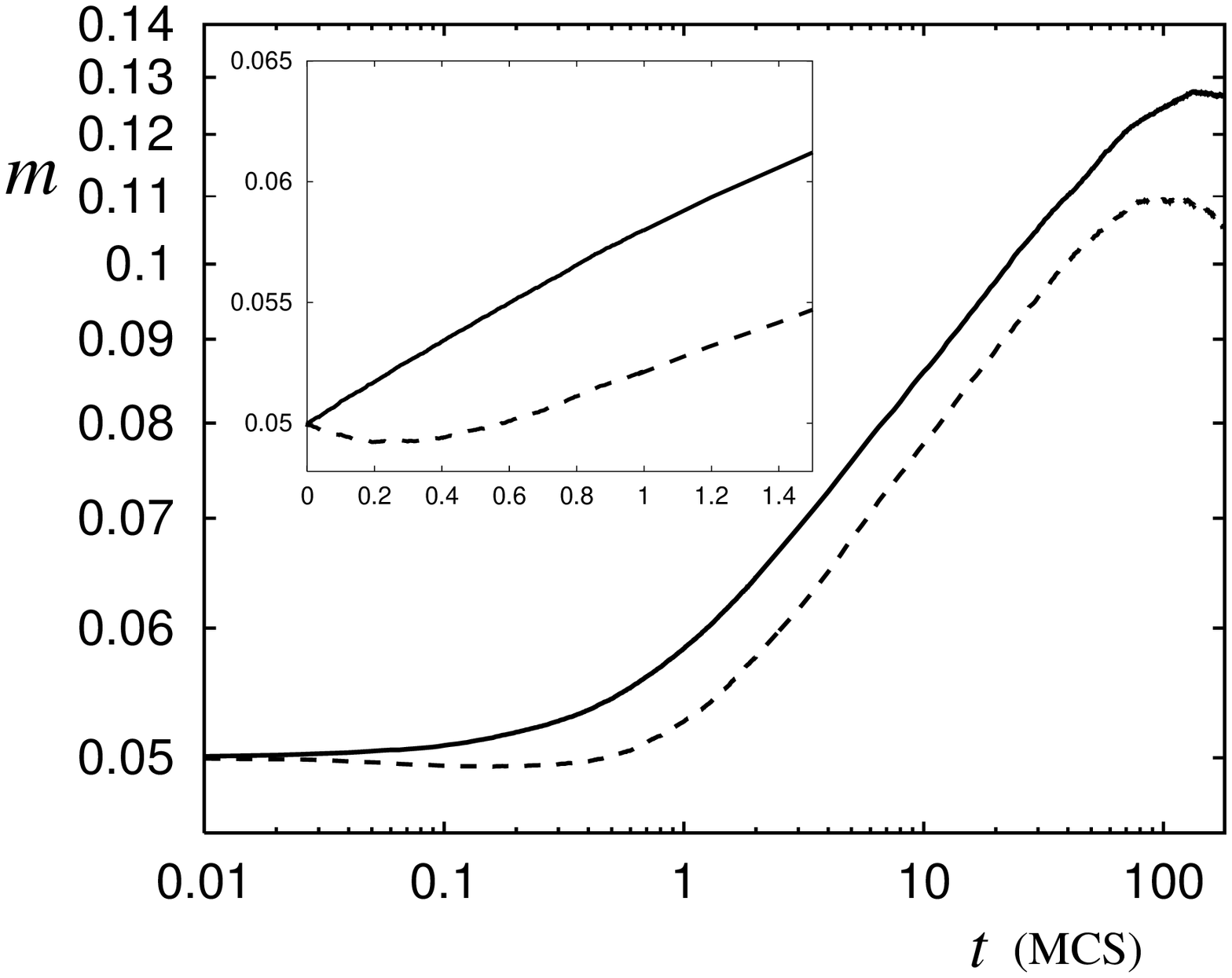}\\[0mm]
{\small {\bf Fig.\,2: } Order-parameter profiles for $L=20$ and $m_0=0.05$
obtained with HB (solid line) and ME algorithm (dashed line) in
double-logarithmic representation. The small diagram inserted shows
the data for small times in double-linear representation.}\\[0.1cm]

Taking into account these results, the
natural question to ask is whether there exists
an algorithm-independent limiting temperature, a
``dynamic MF temperature'', where the
{\it macroscopic} behavior changes from increasing (USTB)
to decaying.
It turns out that for the
HB algorithm this limiting value coincides
with $T^{HB}_l$; the profiles have mostly one extremum.
In order to determine the corresponding
limit for ME, we generated a number of profiles for
temperatures above $T_c$, seeking the one that shows
a saddle point. We detemined the corresponding temperature as
$2.70\,(2)$. This number did neither depend
significantly
on the system size nor on $m_0$.
However, it does not agree with the corresponding value
of the HB dynamics $T^{HB}_l\simeq 3.1$.

Eventually we compared magnetization profiles
in the {\it universal} regime for a system with
$L=40$ and $m_0=0.03$ at the critical point, $T_f=T_c$, for
various combinations of algorithms, lattice types, and updating schemes.
For the square lattice with nearest-neighbor interactions
we combined ME and HB algorithm
with random and sequential updating (4 curves),
for the triangular lattice with nearest-neighbor interactions
we used both algorithms and sequential updating (2 curves),
and for a square lattice with additional next-nearest-neighbor
interactions we used the HB algorithm and sequential updating (1 curve)
\cite{foot3}.

As can be seen already from Fig.\,2,
even though the initial power law is assumed
by both profiles depicted there, the heights and locations
of the maxima depend on the details of the method,
besides the differences for early times. 
However, in all cases studied it turned out to be possible
to map the data onto a single curve for times $t\gtrsim 20$, 
by constant rescalings both axes. 
The result is shown in  Fig.\,3.
On the semi-logarithmic plot (small insertion) the individual
profiles cannot be distinguished. When both axes are 
plotted logarithmically, on the other hand, the short-time
regime is more pronounced, and significant differences for
$t\lesssim 10$ become visible. The pure power law
(solid line above the data) is plotted for comparison.\\[3mm]

\def\epsfsize#1#2{0.6#1}
\hspace*{1cm}\epsfbox{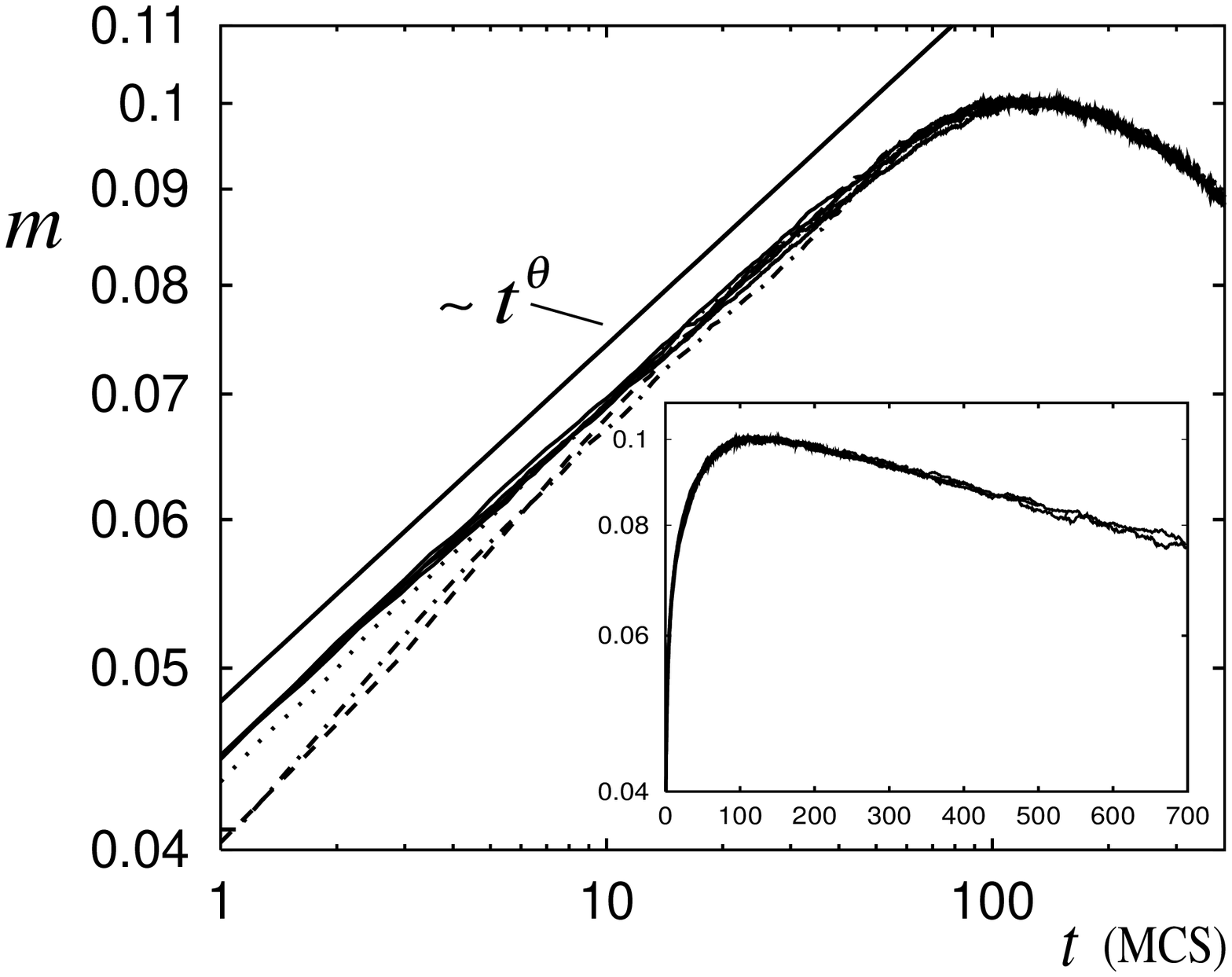}\\[0mm]
{\small {\bf Fig.\,3: } Data collapse of seven magnetization profiles
obtained for $L=40$ and $m_0=0.03$ for different combinations
of lattice types, algorithms, and updating schemes (as described
in the text) in double-logarithmic representation.
The ME results for the square lattice with
sequential (dashed line) and random (dotted line), as well as
for the triangular lattice with sequential update (dashed-dotted line)
are singled out. The small diagram inserted shows the same data in
semi-logarithmic representation. The power law (\ref{power}) with $\theta=0.19$
is plotted for comparison.}\\[0.1cm]

Singled out are the ME curves, for the square
lattice with sequential (dashed) and random (dotted) updating,
and for the triangular lattice with sequential updating (dashed-dotted).
In particular when the ME algorithm is combined with sequential
updating, the power law is assumed only for $t\gtrsim 10$.
This is consistent with the
findings of Okano et al.\cite{okano} for the Potts model, and
can now be interpreted as a
consequence of the anomalous behavior of the
ME algorithm for $t\lesssim 1$.
The HB results assume the power-law form much earlier. 
Most importantly, however, all seven curves show
the USTB as expressed in (\ref{power}) for later times,
$t\gtrsim 10$.
 
In conclusion, we studied the short-time behavior
in relaxational processes after a temperature
quench. We found surprisingly different temporal evolutions with heat-bath
and Metropolis algorithm during early, non-universal stages.
Nevertheless, the characteristic short-time
power law (\ref{power}) turned out to be a rather robust phenomenon, occurring
independently of the algorithm, the lattice type, and the
updating scheme, provided the systems belong to the same
dynamic universality class---the Ising system with short-range
interactions and non-conserved
order parameter
was the example studied above---and the correlation length
has grown substantially larger than microscopic scales.\\[4mm] 
{\small {\bf Acknowledgements}: This work was supported in part by the
Deutsche Forschungsgemeinschaft through
Sonderforschungsbereich 237 ``Unordnung und gro{\ss}e Fluktuationen''.}


\begin{thebibliography} {99}
%
\bibitem{bray} See A. J. Bray, Advances in Physics, {\bf 43}, 357 (1994) for
a recent review on phase-ordering kinetics after quenches to
low temperatures.
%
\bibitem{jans} H.\ K.\ Janssen, B.\ Schaub, and B.\ Schmittmann,
Z.\ Phys.\ B {\bf 73}, 539 (1989).
%
\bibitem{jans2}{H.\ K.\ Janssen, in {\it From Phase Transitions
to Chaos --- Topics in Modern Statistical Physics},
edited by G.\ Gy\"orgyi, I.\ Kondor, L.\ Sasv\'ari, and T.\ Tel
(World Scientific, Singapore, 1992).}
%
\bibitem{huse} D. A. Huse, Phys. Rev. B {\bf 40}, 304 (1989); see
also K. Humayun and A. J. Bray, J. Phys. A: Math. Gen.
{\bf 24}, 1915 (1991).
%
\bibitem{monte} Z.-B. Li, U. Ritschel and B. Zheng, J. Phys. A:
Math. Gen.
{\bf 27}, L837 (1994).
%
\bibitem{liea} Z.-B. Li, L. Sch\"ulke, B. Zheng, Phys. Rev. Lett.
{\bf 74}, 3396 (1995); Phys. Rev. E {\bf 53}, 2940 (1996)
%
\bibitem{sczh} L. Sch\"ulke and B. Zheng, Phys. Lett. A {\bf 204},
295 (1995), {\it ibid.} A {\bf 215}, 81 (1996).
%
\bibitem{okano} K. Okano, L. Sch\"ulke, K. Yamagishi, and
B. Zheng, {\it Universality and Scaling in Short-Time Dynamics},
Siegen preprint (1995).
%
\bibitem{oerd1} K. Oerding and H. K. Janssen, J. Phys. A:
Math. Gen. {\bf 26}, 3369 (1993);  {\it ibid.} {\bf 26}, 5295 (1993).
%
\bibitem{oerd2} H. K. Janssen and K. Oerding, J. Phys. A: Math. Gen. {\bf 27}, 715 (1994).
%
\bibitem{oerd3} K. Oerding, J. Stat. Phys. {\bf 78}, 893 (1995);
J. Phys. A: Math. Gen. {\bf 28}, L639 (1995);
K. Oerding and H. K. Janssen, J. Phys. A {\bf 28}, 4271 (1995);
H. K. Janssen, K. Oerding, and E. Sengespeick , J. Phys. A {\bf 28},
6073 (1995).
%
\bibitem{own} H.\ W.\ Diehl and U.\ Ritschel, J.\ Stat.\ Phys.
{\bf 73}, 1 (1993);
U. Ritschel and H. W. Diehl, Phys. Rev. E
{\bf 51}, 5392 (1995).
%
\bibitem{nucl} U. Ritschel and H. W. Diehl,
Nucl. Phys. B {\bf 464}, 512 (1996).
%
\bibitem{surf} U. Ritschel and P. Czerner, Phys. Rev. Lett. {\bf 75},
3882 (1995); S. N. Majumdar and A. M. Sengupta, 
Phys. Rev. Lett. {\bf 76},   (1996).
%
\bibitem{grass} P. Grassberger, Physica A {\bf 214}, 547 (1995).
%
\bibitem{kothou} P. Czerner and U. Ritschel, Phys. Rev. E {\bf 53},
3333 (1996).
%
\bibitem{zheng} B. Zheng, Phys. Rev. Lett. {\bf 77}, 679 (1996).
%
\bibitem{overshoot} H. Gilh{\o}j, C. Jeppesen, and
O. G. Mouritsen, Phys. Rev. Lett {\bf 75}, 3305 (1995).
%
\bibitem{maju} S. N. Majumdar, A. J. Bray, S. J. Cornell, and C. Sire, 
{\it Global persistence exponent for critical dynamics}, 
cond-mat/9606123.
%
\bibitem{halp}
P.\ C.\ Hohenberg and B.\ I.\ Halperin,
Rev.\ Mod.\ Phys.\ {\bf 49}, 435 (1977).
%
\bibitem{private} H. K. Janssen, private communication.
%
\bibitem{bind} K. Binder and D. W. Heermann, {\it Monte Carlo Simulation
in Statistical Physics}, (Springer, Berlin, 1988) provides
a good introduction in the dynamic interpretation
of Monte Carlo simulations and the different algorithms
used.
%
\bibitem{foot1} For descriptions of the heat-bath algorithm
see B. Derrida and G. Weisbuch, Europhys. Lett. 4, 657 (1987) and
A. M. Mariz, H. J. Herrmann, and L. de Arcangelis,
J. Stat. Phys. {\bf 59}, 1043 (1990).
For the MC simulations reported in this work
the heat-bath algorithm is equivalent to the original Glauber
dynamics, see A. M. Mariz and H. J. Herrmann, J. Phys. A {\bf 22},
L1081 (1989). 
%
\bibitem{foot2} For a detailed discussion of the later stages, especially
the time scales which determine the location of the maximum and
the rate of decay to the equilibrium, we refer to Refs.\,\cite{monte,own}.
%
\bibitem{foot3} For the triangular lattice the exact 
critical temperature is $T_c=3.6409\dots$. For the square lattice
with additional next-nearest-neighbor interactions we chose
(for simplicity) the additional coupling to be equal to $J$
(the nearest-neighbor coupling). We determined
the critical temperature for this system as $T_c\simeq 5.26$.
%
\end{thebibliography}
\end{document}